\documentclass[%
reprint,
showpacs,
bibnotes,
amsmath,amssymb,
aps,
prl,
]{revtex4-1}

\usepackage{graphicx}
\usepackage{dcolumn}
\usepackage{bm}
\usepackage{hyperref}
\usepackage{multirow}
\usepackage{array}
\usepackage{booktabs}
\usepackage{ctable}
\usepackage{upgreek}
\usepackage{epsfig}
\usepackage{mathrsfs}
\usepackage{amssymb}
\usepackage{amsbsy}
\usepackage{color}
\usepackage{cancel}
\usepackage{marginnote}
\newcommand{\bcen}{\begin{center}}
\newcommand{\ecen}{\end{center}}
\newcommand{\btab}{\begin{tabular}}
\newcommand{\etab}{\end{tabular}}
\newcommand{\bdes}{\begin{description}}
\newcommand{\edes}{\end{description}}

\newcommand{\beq}{\begin{equation}}
\newcommand{\eeq}{\end{equation}}
\newcommand{\bea}{\begin{eqnarray}}
\newcommand{\eea}{\end{eqnarray}}

\newcommand{\etal}{{\em et.~al.,~} }
\newcommand{\ibid}{{\em ibid.,~}}
\newcommand{\half}{\frac{1}{2}}
\newcommand{\bary}{\begin{array}}
\newcommand{\eary}{\end{array}}
\newcommand{\benum}{\begin{enumerate}}
\newcommand{\eenum}{\end{enumerate}}
\newcommand{\bitem}{\begin{itemize}}
\newcommand{\eitem}{\end{itemize}}

\newcommand{\bk} { \bm{k} }

\newcommand{\bq} { \bm{q} }

\newcommand{\bK} { \mbox{\boldmath $K$}}

\newcommand{\dou}{\partial}

\newcommand{\D}[1]{\mbox{d}{#1}}

\newcommand{\eqn}[1] {eqn.~(\ref{#1})}

\newcommand{\fig}[1]{fig.~\ref{#1}}
\newcommand{\Fig}[1]{Fig.~\ref{#1}}

\makeatletter

\newcommand{\Rmnum}[1]{\expandafter\@slowromancap\romannumeral #1@}
\makeatother

\newcommand{\myfigwidth}{0.32\paperwidth}

\newcommand{\asm}{a_{sm}}
\newcommand{\as}{a_{s}}

\newcommand{\Ef}{E_F}
\newcommand{\kf}{k_F}

\newcommand{\ederiv}{\dou {\cal E}/\dou(-\as^{-1})}
\usepackage{lineno}

\begin{document}



\title{The Nature and  Properties of a Repulsive Fermi Gas in the ``Upper Branch"}
\author{Vijay B. Shenoy$^1$}
\email{shenoy@physics.iisc.ernet.in}
\author{Tin-Lun~Ho$^2$}
\email{jasontlho@gmail.com}
\affiliation{$^1$Centre for Condensed Matter Theory, Indian Institute of Science, Bangalore 560 012, India}
\affiliation{$^2$Department of Physics, Ohio State University, Columbus, OH 43210}



\date{\today}

\begin{abstract}

We generalize the Nozi\'eres-Schmitt-Rink (NSR) method to study the
repulsive Fermi gas in the absence of molecule formation, i.e., in the
so-called ``upper branch".  We find that the system remains stable
except close to resonance at sufficiently low temperatures. With
increasing scattering length, the energy density of the system attains
a maximum at a positive scattering length before resonance. This is
shown to arise from Pauli blocking which causes the bound states of
fermion pairs of different momenta to disappear at different
scattering lengths. At the point of maximum energy, the
compressibility of the system is substantially reduced, leading to a
sizable uniform density core in a trapped gas. The change in spin
susceptibility with increasing scattering length is moderate and does
not indicate any magnetic instability. These features should also
manifest in Fermi gases with unequal masses and/or spin populations.

\end{abstract}


\maketitle

Since the early days of quantum many-body theory, the Fermi gas with a
repulsive  short range interaction has been used as the primary example
of a Fermi liquid state\cite{Gal}.  The
discovery of BEC-BCS crossover\cite{Jin}, however, shows that the
ground state of this system is a molecular condensate, i.~e., the
Fermi liquid state is metastable.  In the last two years, after the
ref.~\cite{Ketterle} suggested the evidence of Stoner ferromagnetism, there has been increased interest in
the nature of uncondensed Fermi gas (free of molecules) in the
strongly interacting regime. Such systems have been referred to as the
``upper branch" Fermi gas, and the molecular condensate as the ``lower branch".

Theoretical studies have found both ferromagnetic transition as well
as the absence of it.\cite{theory} Though seldom
emphasized, the upper branch Fermi gas in the strongly interacting
regime has been studied by many experimental
groups\cite{expt,Salomon} at higher temperatures with
different densities and trap depths. The key features in atom loss and
energy maximum reported in ref.\cite{Ketterle} also appeared in these
earlier experiments. A puzzling feature is the presence of a range of
scattering length $(\as)$ where the energy derivative is negative
$(\ederiv <0)$ in apparent violation of the
adiabatic relation of Tan\cite{Tan}.  Since the Fermi gas is
unlikely to be ferromagnetic in the temperature regime of these
earlier experiments, it leads to a natural and intriguing question on the
nature of the repulsive gas in the strongly interacting regime.

The key obstacle in theoretical studies of the upper branch Fermi gas
is to find a proper mathematical description of the ``upper branch".
There is no precise formulation of it to the best of our
knowledge. Fortunately, the meaning of upper branch is well defined in
the high temperature regime, as the second virial coefficient $b_{2}$
is made up of a bound state
contribution and an extended (or scattering) state contribution,
$b_{2}=b_{2}^{bd} + b_{2}^{sc}$. The upper branch corresponds to
excluding the Hilbert space of molecules by setting $b_{2}^{bd}=0$. In
addition, any description of the upper branch Fermi gas must also
recover the  results\cite{Gal}
in the weakly interacting limit, as the energies of the molecules are
very far below the continuum and hence can be ignored.

Here we generalize the approach of NSR\cite{NSR}, which we
call the excluded molecular pole approximation (EMPA), to study the
upper branch Fermi gas.  It amounts to excluding the Hilbert space of
molecules in a Guassian fluctuation theory\cite{Randeria},
and obtaining thermodynamics within this truncated space.  This
approach recovers rigorously both the high temperature results and the
results of Galitskii in the weak coupling limit.  Applying this
method to lower temperature and strongly interacting regime, we find
the following: ({\bf I})
On approaching the resonance from the repulsive side at a fixed
temperature $T$, the energy density ${\cal E}$ attains a maximum at a
positive scattering length $(\asm)$ {\em prior} to resonance, as seen
in experiments.\cite{Salomon,Ketterle} The theory also explains the subsequent fall in the energy density with increasing $\as$
(violation of the adiabatic theorem of Tan). ({\bf II}) The
compressibility $\kappa$ attains a minimum at $\asm$ (where ${\cal E}$
is maximum). The small compressibility implies a core of almost
uniform density at the centre of the trap. ({\bf III}) The spin
susceptibility $\chi$ attains a maximum at the location of the energy
maximum, i.~e., at $\asm$; it shows only a moderate
variation over the entire range of $\as$, without any divergence indicative of a 
magnetic instability.

{\em EMPA for the Upper Branch Fermi Gas:} 
Let us first recall that at low fugacity regime \cite{HoMueller}, 
the equation of state is $n(T,\mu)= n_{o}(T, \mu)+ \partial \Delta P/\partial \mu$, where 
$ n_{o}(T, \mu)$ is the density of an ideal gas, 
$\Delta P(T,\mu) = T(\sqrt{2}/\lambda)^3 z^2 b_2 $ is the interaction contribution to the pressure,
$\lambda=\sqrt{\frac{2\pi}{m T}}$ is the thermal wavelength, and $z=e^{\mu/T}$ is the fugacity ($\hbar = k_B = 1$), $\mu$ is the chemical potential.    The second virial coefficient $b_{2}$ is made up of a bound state contribution $b_{2}^{bd}$ and a scattering state contribution $b_2^{sc}$, 
$b_2 = b_{2}^{bd}+ b_2^{sc}$,  
 \begin{equation}
b_{2}^{bd}= e^{|E_{b}|/T}, \,\,\, b_{2}^{sc} = \int_{0}^{\infty} \frac{{\rm d}\omega }{\pi}
 \frac{{\rm d}\eta}{{\rm d}\omega} e^{-\omega/T} ; 
\label{b2-be}\end{equation}
and $-|E_{b}|=-(ma_s^2)^{-1}$ is the energy of the bound state, and $\eta$ is the phase shift. The interaction contribution to equation of state $\Delta n(T,\mu) = n(T, \mu) - n_{o}(T,\mu) $ can therefore be written as 
$\Delta n= \Delta n^{bd} + \Delta n^{sc} $, where 
\begin{equation}
\Delta n^{\alpha}(T, \mu) = \left(\frac{\sqrt{2}}{\lambda}\right)^3  b_{2}^{\alpha} T\frac{\partial z^2}{\partial \mu} ,  \,\,\,\,\,\,  \alpha = bd, \,\, sc \label{highTDn}
\end{equation}

Next we recall that in the NSR approach\cite{NSR}, the interaction contribution to the density
$\Delta n(\mu, T)= n(T,\mu)-n_{o}(T,\mu)$ is 
\begin{equation}
 \Delta n (T,\mu)=  -\frac{1}{\Omega}\sum_{\bq}   \int^{\infty}_{-\infty} \frac{ {\rm d} \omega}{\pi}  n^{}_{B}(\omega) \frac{\partial{\rm arg} M(\omega^{+}, {\bq})}{\partial \mu}.
\label{Dn}  
\end{equation}
where  $\Omega$ is the volume, $n_{o}(T,\mu) = 2 \sum_{\bk} n_{F}^{}(\xi_{\bk})$, 
is the density of a two-component ideal Fermi gas, $n^{}_{F}(\omega) = 1/(e^{\omega/T}+1)$, $\xi_{\bk}= \epsilon_{\bk}-\mu$, $\epsilon_{\bk}=  k^2/2m$, 
$n^{}_{B}(\omega)$ $= 1/(e^{\omega/T}-1)$,  and $M(\omega^{+}, {\bq})$ is negative inverse of the two particle T-matrix in the medium, of the form
\begin{align}
M(\omega^{+}, {\bq}) & = -\frac{1}{4\pi a_{s}}+   L(\omega^{+}, {\bq}) \label{M} \\
 L(\omega^{+}, {\bq}) & = 
\frac{1}{\Omega} \sum_{\bk}\left(  \frac{\gamma ({\bk}; {\bq})}{ \omega^{+} - \xi_{\frac{\bq}{2}+{\bk}} -\xi_{\frac{\bq}{2}-{\bk}} } + \frac{1}{2\epsilon_{\bk}}\right), \label{L} 
\end{align}
  $\gamma({\bk}; {\bq}) = 1- n_{F}(\xi_{\frac{\bq}{2}+{\bk}}) -  n_{F}(\xi_{\frac{\bq}{2}-{\bk}})$ describes Pauli blocking of pair fluctuations.
In the extreme dilute limit, $\gamma({\bk}; {\bq})$ reduces to 1, $-M^{-1}$ to the inverse two-body $T$-matrix, and the phase angle $\zeta(\omega, {\bq})\equiv {\rm arg}M(\omega^{+}, {\bq})$ to the negative of the two-body phase shift $\eta (\omega-\omega(q))$, where 
$\omega(q)=q^2/4m -2\mu$.

\begin{figure}
\centerline{\includegraphics[width=\myfigwidth]{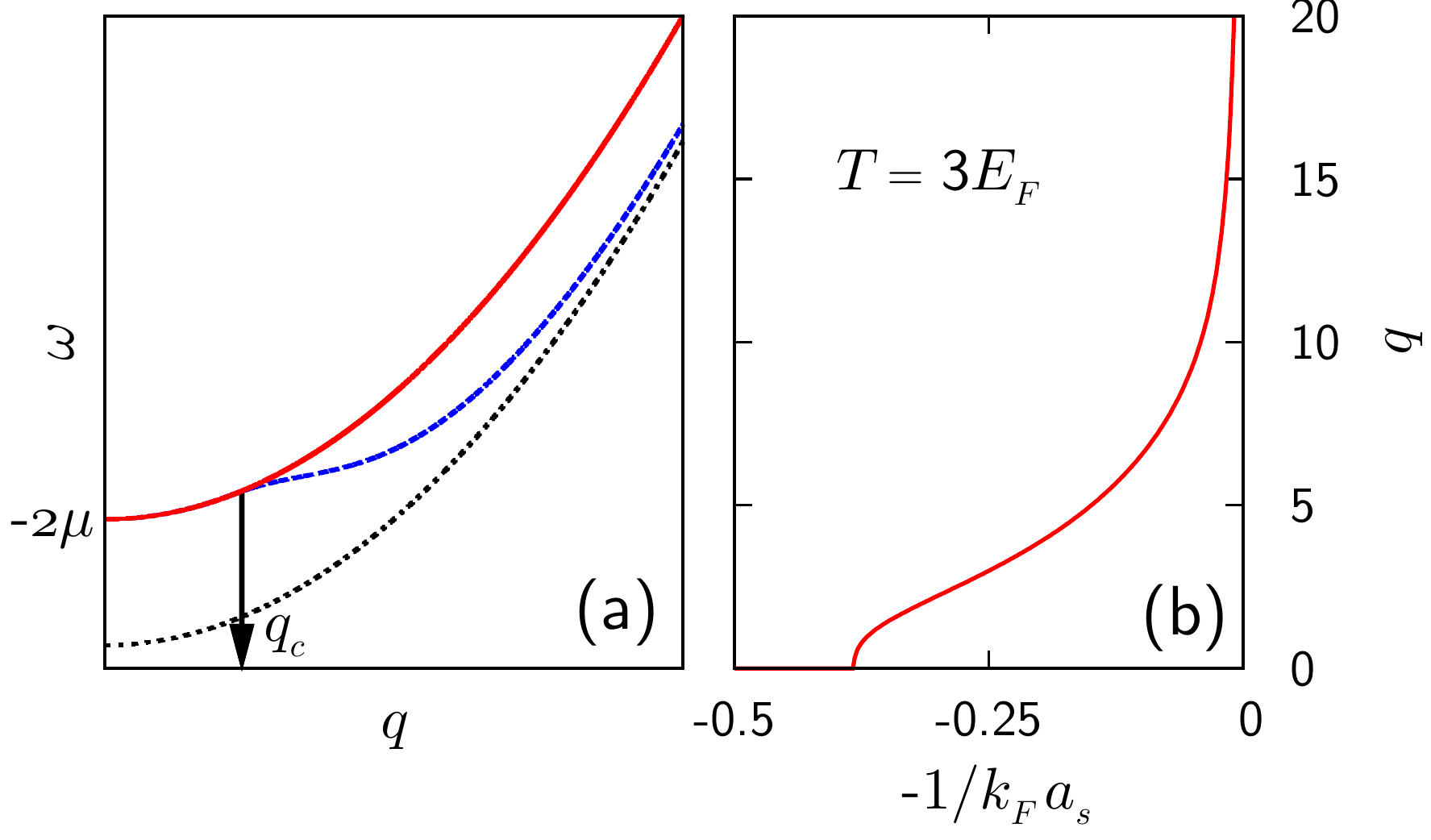}}
\caption{(Color online) {\sf (a)} Position of pole $\omega_{b}(q)$ in $\omega-q$ plane for a given $a_s$:  
The solid curve denotes the curve $\omega(q) = q^2/(4m)-2\mu$.  These pole position $\omega_{b}(q)$ is the solution of the equation 
${\rm Re}M(\omega^{+}, {\bq})=0$ (see \eqn{M}), where Pauli blocking is described by $\gamma({\bk; q})$. 
For $a_s>0$,  the matrix $M(\omega^{+}, {\bq})$ of a  two-body system will have a pole of energy $-|E_{b}|$ below $\omega(q)$, (dotted line).  In a many body system, Pauli blocking will  suppress formation of molecular bound states. The suppression is strongest for pairs with total momentum $q=0$ and is less strong for larger $q$.  As a result the pole position changes to those indicated by the dashed blue curve.   {\sf  (b)} The red curve is the critical scattering length $a_s^{c}(q)$  at $T=3E_F$. (See (B) in {\em Summary of Results}). For a given $a_s$, a fermion pair with total momentum $q$ (referred simply as ``$q$-pair'') can have a bound state only when $a_s <a_s^{c}(q)$, i.e., to the left of the red curve. As $a_s$ increases, such that $a_s$ crosses $a_s^{c}(q)$ from left to right,  a $q$-pair will lose its bound state, and the energy of the scattering state of this pair will jump downward abruptly (see \fig{fig:ejump}). }
\label{fig:PlSchemeQc}
\end{figure}

For a given ${\bq}$, the  value of  $\zeta(\omega, {\bq})$ depends on the location of branch cut and poles of $M^{-1}(\omega^{+}, {\bq})$.  It is clear from Eq.(\ref{M}) and (\ref{L}) that the branch cut is given by 
  $\omega > \omega(q)$. Should $M^{-1}(\omega^{+}, {\bq})$ have a pole, say at $\omega_{b}(q)<\omega(q)$, then we have 
 \begin{align}
\!\!\!\!\!  \omega > \omega(q) , \,\,\,\, \zeta(\omega, {\bq}) = {\rm tan}^{-1}&\left(\frac{ {\rm Im}L(\omega, {\bq})}{ -\frac{1}{4\pi a_{s}} + {\rm Re}L(\omega, {\bq}) }\right), 
\label{I} \\
\!\!\!\!\!  \omega_{b}(q)< \omega < \omega(q), \hspace{0.05in}  \,\,\,\,\,\,   \zeta(\omega, {\bq})& = -\pi,  
\label{II} \\
\!\!\!\!\!  \omega< \omega_{b}(q), \hspace{0.05in}  \,\,\,\,\,\,   \zeta(\omega, {\bq}) &= 0. 
\hspace{0.6in}
\label{III} 
\end{align}
Otherwise,  Eq.(\ref{II}) and (\ref{III}) are replaced by 
\begin{equation}
 \omega< \omega(q), \hspace{0.2in}  \,\,\,\,\,\,   \zeta(\omega, {\bq}) = 0. 
\hspace{0.6in}
\label{IV} \end{equation}

Eq.(\ref{Dn}) can then be written as $ \Delta n^{}(T,\mu) =  \Delta n^{bd}(T,\mu) +  \Delta n^{sc}(T,\mu) $,
\begin{align}
 \Delta n^{bd}(T,\mu) &= -\frac{1}{\Omega}\sum_{\bq}  n^{}_{B}(\omega_{b}(q))  \frac{\partial \omega_{b}(q)}{\partial \mu} \label{Dn1}  \\
 \Delta n^{sc}(T,\mu) & = -\frac{1}{\Omega}\sum_{\bq}   
 \int^{\infty}_{\omega(q)}
 \frac{ {\rm d} \omega}{\pi}  n^{}_{B}(\omega) \frac{\partial\zeta(\omega, {\bq})}{\partial \mu}. 
\label{Dn2}  
\end{align} 
That we use the same superscript in Eq.(\ref{Dn1}) and (\ref{Dn2}) as in the high temperature case is because they reduce to Eq.(\ref{highTDn}) in the  low fugacity regime. 
Thus, by continuity, the extension of the upper branch Fermi gas 
 to lower temperature is to 
{\em exclude the contribution from the molecular bound pole term ( Eq.(\ref{Dn1})) from $\Delta n(T,\mu)$}. Hence the name EMPA. The equation of state within 
EMPA is then
\begin{equation}
n(T, \mu) = n_{o}(T, \mu) + \Delta n^{sc}(T, \mu).
\label{eos} \end{equation}
Considering a system with fixed density $n$ (which defines $E_F$ and  $k_F$) and inverting the relation $n=n(T,\mu)$ to obtain  $\mu=\mu(n,T)$, one can obtain all thermodynamic potentials as a function of $n$ and $T$.\cite{thermo} It is also straightforward to describe spin polarized states in the upper branch by  including a magnetic field $h$.

Note that Eq.(\ref{eos}) involves only integrating over the area  $\omega>\omega(q)$ (i.~e., above the solid curve in \Fig{fig:PlSchemeQc}(a)) with an integrand given  explicitly by Eq.(\ref{I}). There is no need to obtain the pole structure as far as evaluating Eq.(\ref{eos}) is concerned. There is, however,  a close connection between the interaction energy of scattering state and the presence of a pole. Understanding the distribution of poles in the $\omega$-$q$ plane is, therefore, essential for the elucidation of the results to be presented below.

\begin{figure}
\includegraphics[width=\myfigwidth]{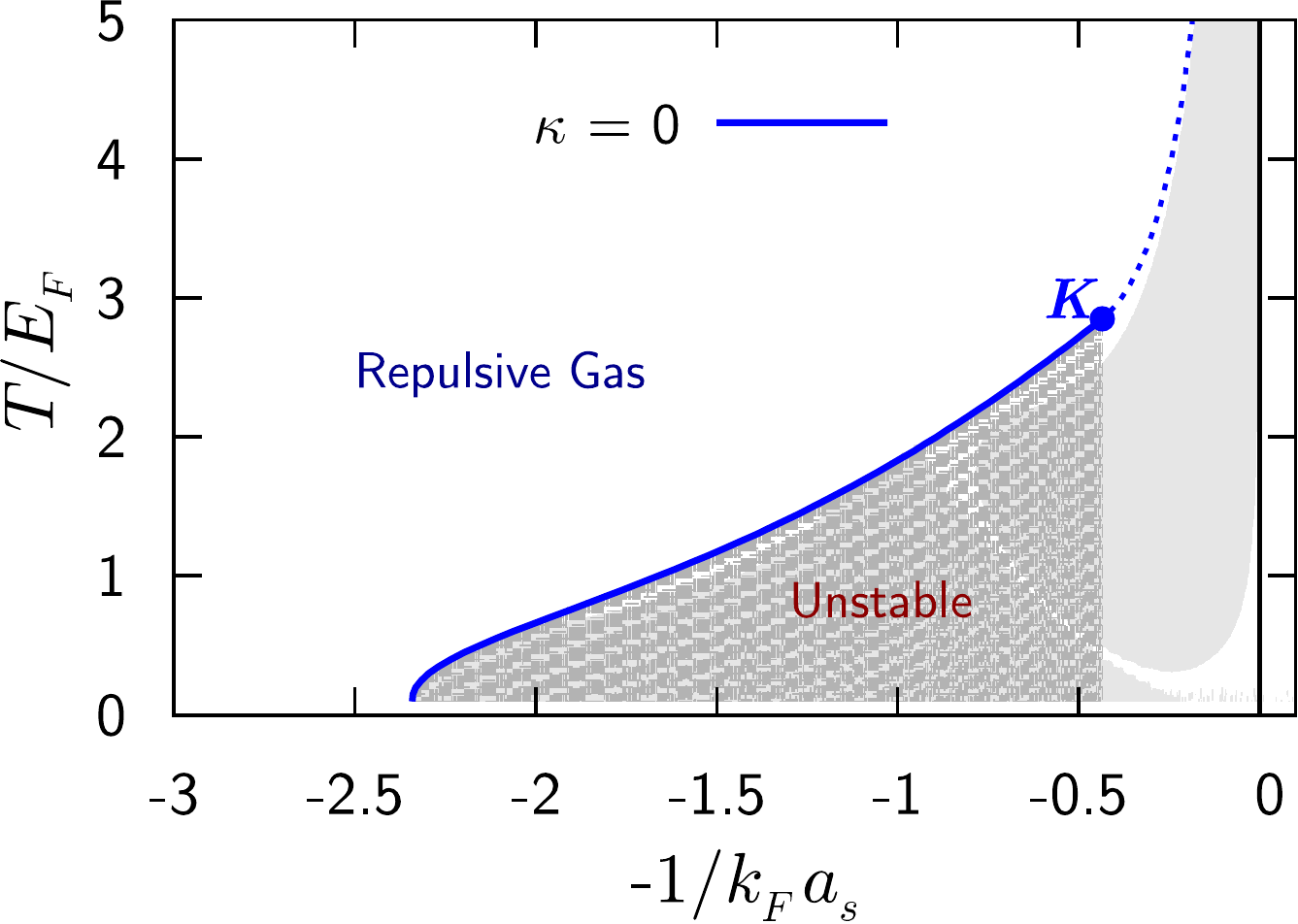}
\caption{(Color online) Upper branch ``phase diagram''.  The point $\bK$ corresponds to $(-1/\kf\as = -0.435, T = 2.85 \Ef)$. The solid blue line ending at $\bK$ is a locus of states with a vanishing compressibility. The dashed curve starting at $\bK$ shows $\asm$ (see text).  The region hatched in dark grey is mechanically unstable. Tan's adiabatic theorem is violated in the region shaded in light grey.}
\label{fig:UpperPD}
\end{figure}

\noindent{\em Summary of Results:}

\noindent  {\em (A) Phase diagram:} \Fig{fig:UpperPD} displays the ``phase diagram'' of the  upper-branch Fermi gas. All regions except the region in dark grey are stable ($\kappa,\chi>0$). The solid line that ends at $\bK$ is the boundary of vanishing compressibility  $\kappa=0$.   
The dashed line above $\bK$ describes $\asm$ where the energy attains a maximum at a fixed temperature. Across this line, $\mu$, $\kappa$,  $\chi$, and energy density ${\cal E}$ are continuous but their slope undergoes sharp changes.  These discontinuous slopes, however,  may disappear if beyond Gaussian fluctuations are included. Crossing the solid line below  the  point ${\bK}$, the quantitites  $\mu$, $P$, $\kappa$,  $\chi$, and ${\cal E}$ undergo  discontinuous changes; the system is mechanically unstable.
 The white and light grey regions correspond to regimes with $\ederiv > 0$ and $\ederiv < 0$ respectively. In the light grey region, Tan's adiabatic theorem is not applicable (see below).

\noindent {\em (B) Energy density ${\cal E}$:} \Fig{fig:highT}(a) shows the behavior of energy density ${\cal E}$ as a function of $k_{F}a_s$ at $T=3T_F$.  It exhibits a maximum at $k_{F}\asm = 2.61$, (which  falls on the dashed  line in \Fig{fig:UpperPD}). 
Such a maximum feature is consistent with the early observation by Salomon's group\cite{Salomon} at high temperatures, as well as in ref.\cite{Ketterle} at lower temperatures. The maximum behavior implies that there is a region of $k_{F}a_{s}$ (the light grey region in \Fig{fig:UpperPD}) where the 
the adiabatic theorem, $\ederiv > 0$ is violated.
  The resolution is that the relation between $\ederiv$ of the scattering state and the contact density is ill-defined at the scattering length where a molecular bound state disappears.

\begin{figure}
\centerline{\includegraphics[width=\myfigwidth]{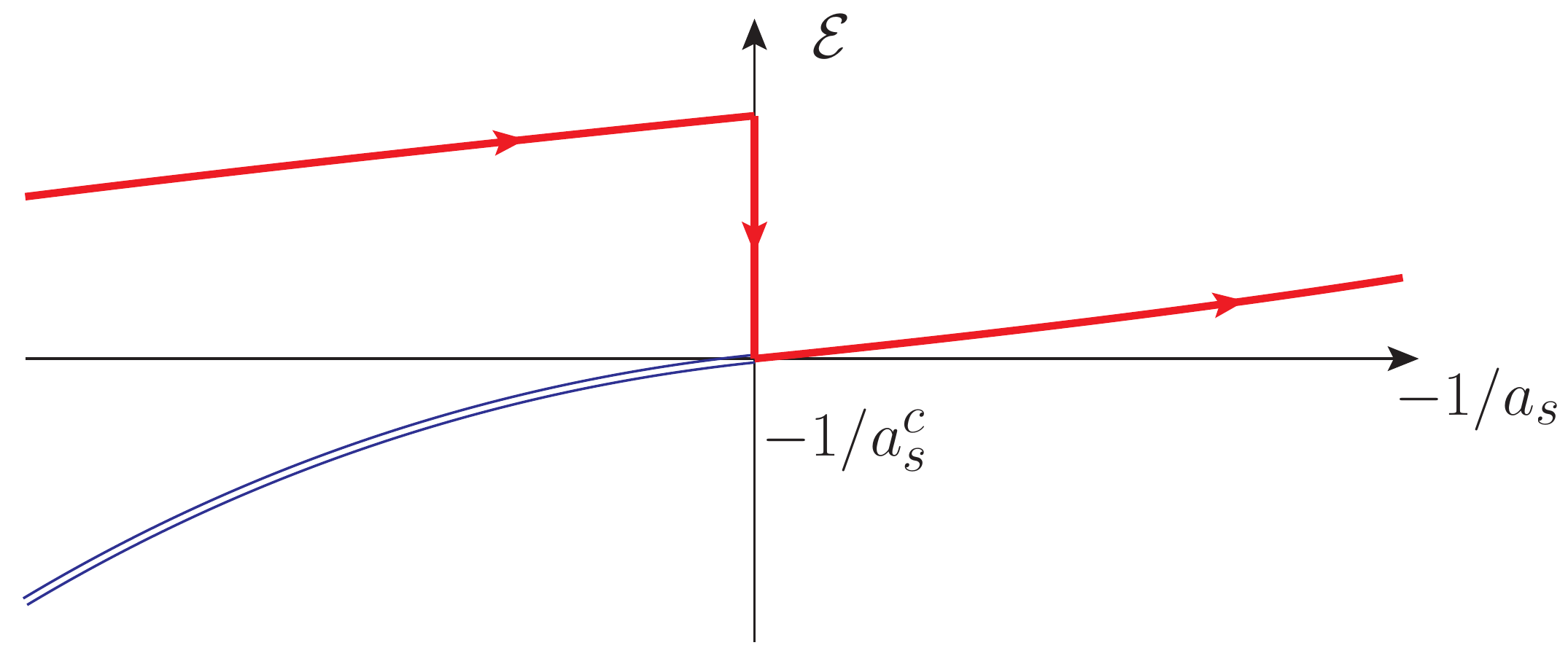}}
\caption{(Color online) The discontinuous change of the energy of the scattering state (solid line) of a two body system up on the disappearance of the molecular bound state (double line). \cite{HoMueller} Similar phenomena occur in a $q$-dependent fashion in the many body setting.
}
\label{fig:ejump}
\end{figure}

This is best seen in the two-body case (see \Fig{fig:ejump}), where the energy of the scattering state of a fermion pair  with total momentum ${\bq}$ (referred to as  ``${\bq}$-pairs'')  jumps suddenly downward when of $a_s$ passes a critical value ($(a_{s}^{c})^{-1}=0$ in this case) at which the molecular bound state on the side $a_{s}<(a_{s}^{c})^{-1}$ disappears. 
In the many-body case, due to Pauli blocking, $(\gamma({\bk}; {\bq})\neq 1)$, different ${\bq}$-pairs will form bound states at different critical scattering length 
$a_{s}^{c}(q)$ (which is the lowest value of $a_{s}$ such that the equation Re$M(\omega, {\bq})=0$ has a solution).  Since Pauli blocking effect is strongest for the ${\bq}=0$ molecular bound state (\Fig{fig:PlSchemeQc}(a)), and is less significant as $q$ increases, $(a_{s}^{c}(q))^{-1}$ is largest at $q=0$ and decreases monotonically as $q$ increases. The behavior of $a_{s}^{c}$ is shown in \Fig{fig:PlSchemeQc}(b), and 
$a_{sm}\equiv a_{s}^{c}(q=0)$.

\begin{figure}
\includegraphics[width=\myfigwidth]{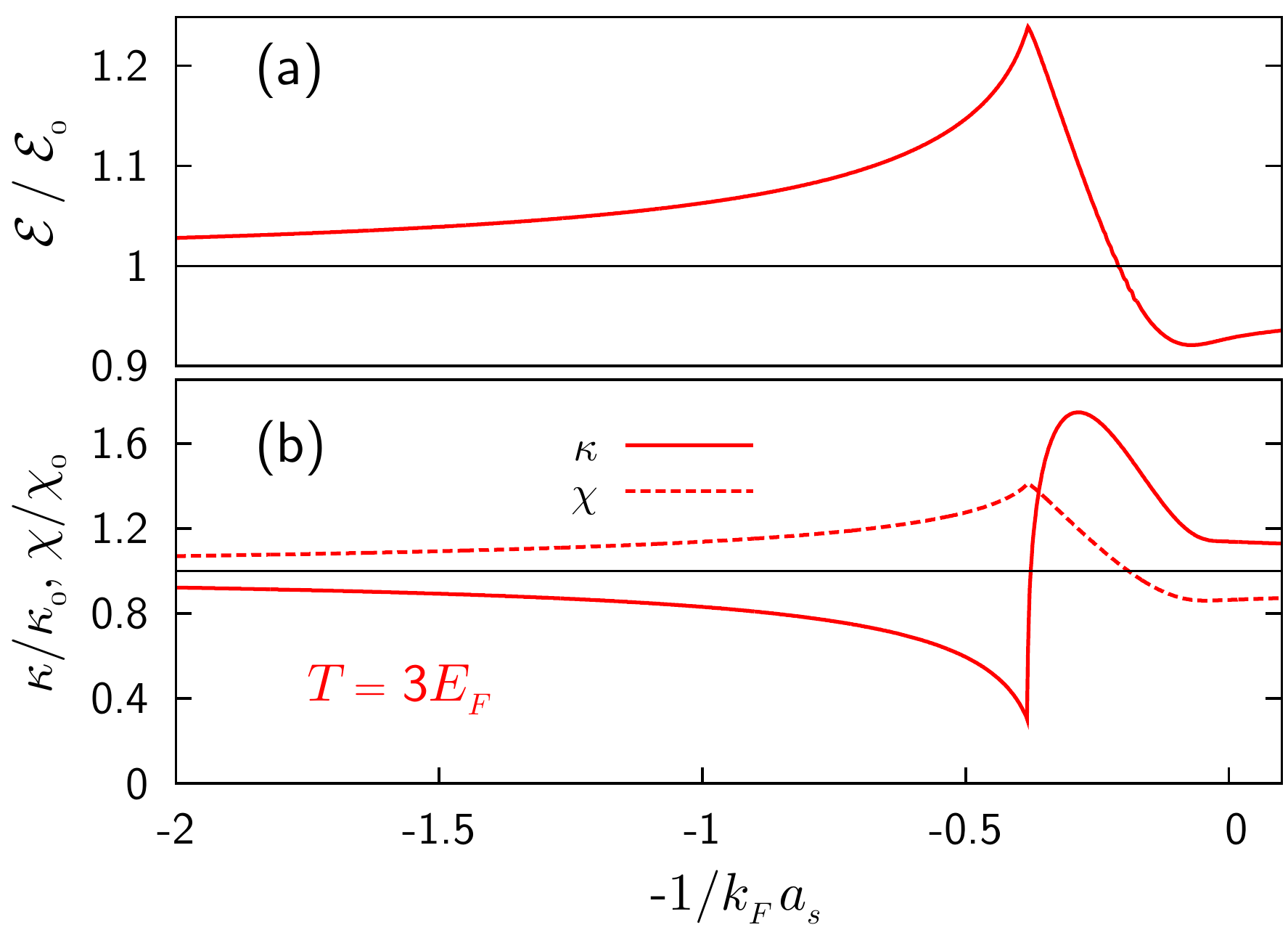}
\caption{(Color online) (a) Energy density $({\cal E})$ and (b) compressibility ($\kappa$)
 and susceptibility ($\chi$) as a function of the scattering length at $T=3E_F$. All quantities are measured in units of their respective values of the non-interacting gas (indicated by the subscript $o$) at the same temperature.
}
\label{fig:highT}
\end{figure}

That $\ederiv <0$ for upper branch Fermi gas sufficiently close to 
resonance is now clear. As $a_s$ passes through  $a_{s}^{c}(q)$ from below, the molecular bound state of a $\bq$-pair disappears because of Pauli blocking. Up on this  disappearance,  the energy of the scattering states of this pair suddenly jumps down, thereby causing the energy to decrease. As $a_s$ continues to increase,  $\bq$-pairs with successively higher total momentum $q$ lose their bound states,  inducing a successive downward jump in the energies of the scattering states of these pairs, and hence a negative derivative $\ederiv <0$. 
Note that since $\asm$ is determined only by Pauli blocking, it should be a universal function of $T$ and $n$  i.e., $k_{F} \asm = f(T/E_{F})$, where $f$ is a dimensionless function (dashed line in \Fig{fig:UpperPD}). 


Our explanation above might lead one to think that the energy decreasing process will cease when 
no more $\bq$-pairs lose their bound states, which occurs at $a_s=\infty$. What actually happens, however, is that the minimum of ${\cal E}$ as $a_s$ increases beyond $a_{sm}$ (which signifies the ceasing of energy decrease) 
occurs at a scattering length prior to resonance. The reason is that in order to have an energy decrease caused by the scattering state of a $\bq$-pair, this pair state has to be occupied.  At lower temperatures, the probability of occupation of such pair states is low especially for those pairs with high $\bq$, thereby causing the energy decrease to cease at an $(a_s)_{min}$  prior to resonance. As $T$ increases, $((a_{s})_{min})^{-1}$ approaches 0.

\begin{figure}
\centerline{\includegraphics[width=\myfigwidth]{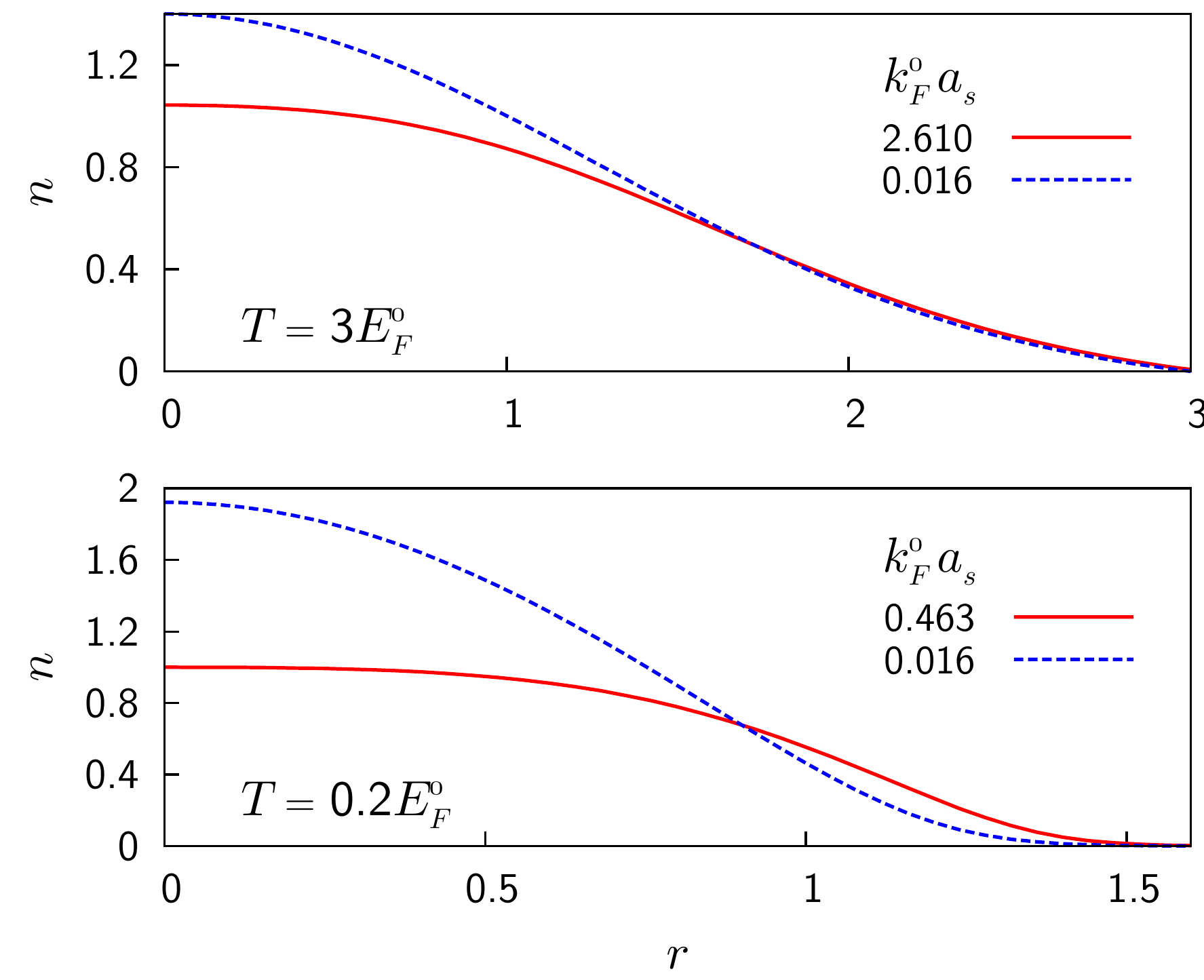}}
\caption{(Color online) Comparison of densities of  strongly interacting and weakly interacting gases in a spherical trap.  $E_F^o$ corresponds to the density at the trap centre.  For each temperature, the number of atoms in both the strong and weak cases is the same. A ``flat-top'' density profile is evident in the strong case, and becomes more pronounced at lower temperatures. The radius $r$ is in units of $\sqrt{(\frac{2E_F^0}{m \omega_t} )}$ ($\omega_t$ - trap frequency).}
\label{fig:TrapDens}
\end{figure}

\noindent {\em (C) Compressibility $\kappa$}:
As $\as$ increases, a
repulsive Fermi gas is expected to become less compressible. For
temperatures above that of point $\bK$ in \Fig{fig:UpperPD},
$\kappa$ attains a minimum at $\as=\asm$ (see \Fig{fig:highT}(b)).
Our calculation shows, for temperatures lower than that of $\bK$,  $\kappa
\rightarrow 0$ as one approaches the solid line (\Fig{fig:UpperPD})
from the left. The system behaves like a hard core Fermi gas with a core size close
to inter-particle spacing. There is, however,  an important difference between a
hard core Fermi gas with core size equal to $a_{s}\approx k_F^{-1}$ and
the actual atomic Fermi gas. In the former case, the effective range
is also of order $k_F$, whereas the effective range in atomic gases is
much less than the inter-particle spacing, independent of the value of
$a_s$.  The diminished compressibility has a dramatic effect
on the density profile. This leads to clouds with little variation of
density at the centre, an effect that becomes more pronounced at lower
temperatures (see \Fig{fig:TrapDens}).

\noindent {\em (D) Spin susceptibility $\chi$:}  \Fig{fig:highT}(b) also shows the spin susceptibility $\chi$ at $T=3T_F$.   Note that $\chi$ changes by at most 40 percent for the entire $k_{F}a_s$ range, and only moderately in the experimentally relevant range $0.5< k_{F}a_s<2$. We do not see a diverging susceptibility indicative of a magnetic transition.

Although our discussions focused on the equal-mass spin-$\half$ Fermi
gas, the properties enumerated arise mainly from Pauli blocking. These
features should, therefore, be generic to other upper branch Fermi
gases, such as those with unequal masses or unequal spin populations.

VBS thanks DAE (SRC-grant) and DST (Ramanujan-grant) for support.
T-LH is supported by NSF Grant DMR-0907366 and by DARPA under the Army
Research Office Grant Nos.~W911NF-07-1-0464, W911NF0710576, and the
Tsinghua University Initiative Scientific Research Program.  This
paper was completed during the INT Workshop on Fermions From Cold Atom
to Neutron Star in May 2011.


\begin{thebibliography}{99}
\bibitem{Gal} V.~M.~Galitskii, Sov.~Phys.-JETP {\bf 7}, 104 (1958).  
\bibitem{Jin} C.~A.~Regal, M.~Greiner, and D.~S.~Jin, Phys. Rev.~Lett.~{\bf 92}, 040403 (2004)
\bibitem{Ketterle}  G.-B.~Jo \etal Science, {\bf 325}, 1521-1524 (2009).
\bibitem{theory} T.~Sogo and H.~Yabu, Phys.~Rev.~A  \textbf{66}, 043611 (2002),
T.~Maruyana and G. Bertsch, \ibid  \textbf{73}, 013610 (2002),
R.~A.~Duine and A.~H.~MacDonald,  Phys.~Rev.~Lett. \textbf{95}, 230403 (2005),
S.~Pilati, \etal \ibid  \textbf{105}, 030405 (2010), 
S.-Y.~Chang, M.~Randeria, and N.~Trivedi, Proc.~Nat.~Acad.~Sci., \textbf{108}, 51 (2011),
S.~Q.~Zhou, D.~M.~Ceperley and S.~Zhang, arXiv:1103.3534,
C.-C.~Chang, S.~Zhang, D.~M.~Ceperley, Phys.~Rev.~A {\bf 82}, 061603(R) (2010),
L.~J.~Le Blanc, \etal \ibid \textbf{80}, 013607 (2009),
G.~J.~Conduit and B. Simons, Phys.~Rev.~Lett.~\textbf{103},200403 (2009),
H.~Heiselberg, arXiv:1012.4569v1,
H.~Zhai, Phys.~Rev.~A \textbf{80} 051605(R) 2009,
X.~Cui and H.~Zhai, \ibid {\bf 81}, 041602(R) (2010),  
D.~Pekker, \etal Phys.~Rev.~Lett.~{\bf 106} 050402 (2011). 
  
\bibitem{expt}  K.~Dieckmann, \etal Phys.~Rev.~Lett.~{\bf 89}, 203201 (2002), C.~A.~Regal, \etal \ibid {\bf 92}, 083201 (2004). S.~Jochim, \etal \ibid {\bf 91}, 240042 (2003); see also S.~Jochim, dissertation, Bose-Einstein Condensation of Molecules, University of Innsbruck, 2004.

\bibitem{Salomon}   T.~Bourdel, \etal Phys.~Rev.~Lett. {\bf 91}, 020402 (2003).
\bibitem{Tan}  S.~Tan, Ann.~Phys. {\bf 323}, 2952 (2008); {\bf 323}, 2971 (2008) and {\bf 323}, 2987 (2008).
\bibitem{NSR}   P.~Nozi\'eres and S.~Schmitt-Rink, J.~Low~Temp.~Phys.~{\bf 59}, 195 (1985).

\bibitem{Randeria}  C.~A.~R.~S\`a de Melo, M.~Randeria, and J.~R.~Engelbrecht, Phys.~Rev.~Lett. {\bf 71}, 3202 (1993).


\bibitem{HoMueller}T.-L.~Ho and E.~J.~Mueller, Phys.~Rev.~Lett.~{\bf 92}, 160404 (2004) and references therein.

\bibitem{thermo} From the density, pressure $P$, compressibility $\kappa$, entropy density $s$, and total energy ${\cal E}$ can be calculated as  $P(T,\mu)=\int^{\mu}_{-\infty} \D{\mu'} n(T,\mu')$,  $\kappa(T,\mu)= \partial n(T,\mu)/\partial \mu$, $s(T, \mu) =  \partial P/\partial T$,  ${\cal E}= Ts - P + \mu n$. Spin density and susceptibility can be similarly calculated by introducing a magnetic field $h$.


\end{thebibliography}
\end{document}